\documentclass[sigconf]{acmart}

\AtBeginDocument{%
  }

\copyrightyear{2026}
\acmYear{2026}
\setcopyright{cc}
\setcctype{by}
\acmConference[RecSys '26]{20th ACM Conference on Recommender Systems}{September 27-October 02, 2026}{Minneapolis, MN, USA}
\acmBooktitle{20th ACM Conference on Recommender Systems (RecSys '26), September 27-October 02, 2026, Minneapolis, MN, USA}
\acmDOI{10.1145/3773078.3831867}
\acmISBN{979-8-4007-2284-4/2026/09}

\begin{document}


\title[Multi-Objective Ranking for Live Streaming]{Multi-Objective Ranking for Live-Streaming: Balancing Fresh and Delayed Signals with Segment-Aware Targeting}

\author{Xiaoyi Gu}
\email{guixiaoy@twitch.tv}
\affiliation{%
    \institution{Twitch Interactive}
    \city{San Francisco}
    \state{California}
\country{USA}}

\author{Julia Tavares}
\email{juliatav@twitch.tv}
\affiliation{%
    \institution{Twitch Interactive}
    \city{San Francisco}
    \state{California}
\country{USA}}

\author{Eder Santana}
\email{edesanta@twitch.tv}
\affiliation{%
    \institution{Twitch Interactive}
    \city{San Francisco}
    \state{California}
\country{USA}}

\author{Carlos Mendoza-Cardenas}
\email{gcarlmen@twitch.tv}
\affiliation{%
    \institution{Twitch Interactive}
    \city{San Francisco}
    \state{California}
\country{USA}}

\author{Nikita Mishra}
\authornote{Work performed while at Twitch Interactive.}
\email{mishniki@amazon.com}
\affiliation{%
    \institution{Amazon Prime Video}
    \city{Sunnyvale}
    \state{California}
\country{USA}}

\author{Saad Ali}
\email{sali@twitch.tv}
\affiliation{%
    \institution{Twitch Interactive}
    \city{San Francisco}
    \state{California}
\country{USA}}


\begin{abstract}  
  One of the most challenging problems entertainment live-streaming services face in recommendation systems is that user behaviors are sparse and delayed, and interaction data exhibits bias for different user segments. Unlike e-commerce applications where user actions follow linear sequences, live-streaming viewers engage in multiple concurrent behaviors of watching, chatting, following, and spending, each occurring with varying delays. We address these challenges through three key contributions: 1) a delayed window approach that extends feedback collection beyond immediate responses, 2) a multi-model architecture that combines fresh and delayed signals, and a segment-aware targeting module that optimizes ranking scores differently across user lifecycle stages, and 3) Multi-gate Mixture-of-Experts (MMoE) integration that jointly models correlated targets while reducing model parameters by 41.9\% compared to independent models. Online A/B testing demonstrates significant improvements, including a +0.09\% increase in Daily Active Viewers (DAV), generating millions more annual active viewer days, and +0.56\% increase in highly engaged viewers' capped Average Revenue Per User (ARPU). Viewer-segment targeting achieved an additional +0.15\% DAV improvement for newer and less engaged viewers, while MMoE enhancement added +0.08\% overall DAV and +0.27\% new follows. The proposed system processes ranking requests with low latency, providing a scalable approach for balancing multiple business objectives across diverse user populations. In addition, we tested the multi-model architecture on the Twitch mobile live feed and achieved a +1.12\% increase in positive user-channel interactions (clicks, follows, and likes), demonstrating applicability beyond the primary use case.
\end{abstract}




\maketitle

\section{Introduction}
\label{sec:introduction}
A live-streaming service like Twitch is a two-sided marketplace where viewers and streamers interact through multiple mechanisms, including watching content, chatting, following channels, and participating in commerce activities. For sustainable business growth, it is important for services to satisfy the needs of all stakeholders in this ecosystem: viewers seek to both discover new content and form bonds with existing communities; streamers aim to grow audiences and monetize their channels. Therefore, the service's personalized recommendation system needs to balance immediate engagement with long-term retention and revenue goals.

Multi-objective optimization (MOO) in recommendation systems has advanced significantly through multi-task learning architectures that enable flexible knowledge sharing across objectives. Notable approaches include Multi-gate Mixture-of-Experts (MMoE) \citep{ma2018modeling} and subsequent architectures \citep{ma2019snr, tang2020progressive, wang2025home}, which have demonstrated success in large-scale industrial applications \citep{zhao2019recommending}. To balance competing objectives, researchers have explored gradient-based methods \citep{yu2020gradient, bai2024gradcraft} and Pareto optimization strategies \citep{lin2019pareto, xie2021personalized, agah2025pareto}.

While these MOO approaches have proven effective for e-commerce and video-on-demand (VOD) applications, live-streaming services present unique challenges that require specialized solutions. In particular, we identify three core challenges that are amplified in live-streaming where viewers engage through multiple concurrent behaviors of watching, chatting, following, and spending.

The first is \textbf{target sparsity}: high-value user actions such as follows, subscriptions, and purchases occur at rates orders of magnitude lower than primary engagement signals like clicks and short views, making them difficult to model effectively. In live-streaming, this challenge is especially pronounced --- our analysis of millions of impressions from 1 million viewers over 7 days showed a progression of decreasing action density: short watch rates were approximately 2.5x lower than click-through rates, longer watch rates 3.4x lower, chat rates approximately 20x lower, and follow rates nearly 90x lower, with monetary transactions occurring at even lower frequencies.

Target sparsity is compounded by \textbf{delayed feedback}, where user actions may occur hours or days after initial exposure rather than immediately. This creates a fundamental tension: labeling non-responses as negatives too early injects noise, while waiting too long makes training data stale. In live-streaming, a viewer might watch a recommended channel multiple times over several weeks before deciding to follow, and may only subscribe after establishing a stronger bond through repeated viewership.

The third challenge is \textbf{user segment bias}: highly engaged users dominate training data, biasing models away from newer or less active users who represent critical growth potential. This is compounded by differing optimization needs across user lifecycle stages: less engaged users benefit from recommendations focused on immediate engagement to encourage repeated visits, while highly engaged users benefit more from deeper engagement and monetization actions.

To address these challenges, we present a multi-objective optimization framework for a large-scale live-streaming service. We use Twitch's recommendation system as our case study, extending a single engagement-focused ranking algorithm \citep{chen2022weighing} to a multi-objective framework that jointly optimizes engagement, retention, and monetization. We validate our approach through comprehensive offline and online A/B experiments, and further demonstrate generalization to Twitch's feed ranking model.

Our main contributions are as follows:

\begin{itemize}

\item We show that separating fresh and delayed signals is a strong driver of improvement for live-streaming recommendation, and operationalize this through a multi-model architecture combining Fresh Signal Models (FSM) for immediate engagement with Delayed Signal Models (DSM) that aggregate sparse actions over a 14-day delayed window. We introduce a viewer segment targeting (VST) module applying different optimization across user lifecycle stages at inference time.

\item We demonstrate that jointly modeling deeper engagement and delayed actions in MMoE while keeping shallow engagement as an independent FSM improves performance while reducing model parameters by 41.9\%.

\item We validate through staged online A/B on Twitch's recommendation system serving millions of daily viewers. The multi-model with delayed window improved DAV\footnote{DAV: Daily Active Viewers} by +0.09\% and highly engaged viewers' capped ARPU\footnote{ARPU: Average Revenue Per User} by +0.56\%. The VST module further boosted less engaged viewers' DAV by +0.15\%. The MMoE enhancement added +0.08\% overall DAV and +0.27\% new follows\footnote{DAV and ARPU are top-line challenging product metrics; improvements of this magnitude represent significant business impact at Twitch's scale.}. We achieve these improvements while maintaining system latency under 110ms p99 at scale.

\end{itemize}

\section{Related Work}
\label{sec:related_work}

\subsection{Multi-Objective Optimization}
\label{sec:related_work_moo}
Multi-objective optimization (MOO) in recommender systems has advanced significantly through multi-task learning frameworks that enable joint optimization of multiple, often conflicting objectives \citep{standley2020tasks}. Early approaches employed shared-bottom architectures \citep{ruder2017overview} where all tasks share a common representation. Multi-gate Mixture-of-Experts (MMoE) \citep{ma2018modeling} advanced this paradigm by introducing task-specific gating mechanisms over shared expert networks, enabling flexible knowledge sharing across tasks. This architecture has demonstrated success in large-scale systems such as YouTube recommendations \citep{zhao2019recommending}. Subsequent work has explored more sophisticated architectures including SNR \citep{ma2019snr} that modularized the shared hidden layers, PLE \citep{tang2020progressive} that separates the task-common and task-specific parameters explicitly, MSSM \citep{ding2021mssm} that learns features selectively, and HoME \citep{wang2025home} with a hierarchy meta experts structure. For sequential user behaviors, ESMM \citep{ma2018entire} models dependencies between objectives in e-commerce conversion funnels, where actions follow a defined order. AITM \citep{xi2021modeling} adaptively learns sequential dependence among multi-step conversions in display advertising.

Researchers have also explored gradient-based methods \citep{yu2020gradient, bai2024gradcraft}, Pareto optimization \citep{lin2019pareto, agah2025pareto, nikolakaki2025semorec}, policy learning \citep{jeunen2024multi}, efficient adaptation \citep{shen2025paragon}, and objective ensembling and alignment \citep{cao2025pantheon, xia2026harmonrank, wang2026onelive} to balance competing or correlated objectives.

While these MOO approaches have proven effective for industrial applications such as e-commerce and VOD applications, they do not adequately address the unique challenges of live-streaming environments, particularly multiple concurrent delayed feedback and viewer segment bias, which we tackle in this work.

\subsection{Delayed Feedback and Target Sparsity}
\label{sec:related_work_delayed_feedback_sparsity}
In live-streaming recommendation systems, unobserved interactions are not necessarily negative \citep{chen2023bias}. The delayed feedback challenge is compounded by target sparsity, particularly for high-value events such as follows, subscriptions, and purchases that occur at extremely low frequencies, which is also observed in advertising conversion scenarios \citep{agah2025pareto, dishi2025practical, liu2025multi}. Research in display advertising has addressed delayed feedback through probabilistic delay models \citep{chapelle2014modeling}, nonparametric approaches \citep{yoshikawa2018nonparametric}, loss function corrections \citep{ktena2019addressing}, multi-window methods \citep{gao2022multi}, and auxiliary correction models \citep{wang2023unbiased}.

However, most existing methods primarily address sequential conversion scenarios where a single target action (like purchase) follows an initial interaction (like click). They do not address the unique challenges of live-streaming services, where multiple viewer actions (watching, chatting, following) can occur independently after content consumption and require joint optimization. Meanwhile, sparse actions like subscriptions and purchases have insufficient positive labels without extended observation windows of days, making them challenging to capture within short windows \citep{liang2024ensure}.

\subsection{Viewer Segment Bias}
\label{sec:related_work_viewer_segment_bias}
While multi-objective optimization has advanced recommendation systems, existing work rarely addresses how to handle training data imbalance across user lifecycle stages. Users at different engagement levels contribute disproportionately to training data, and meanwhile require different optimization strategies within a unified multi-objective framework.

Bias in recommendation systems has been extensively studied \citep{schnabel2016recommendations, chen2023bias}, but much research focuses on item-level or algorithmic biases \citep{ai2018unbiased, schnabel2020debiasing} rather than segment imbalances. \citet{fu2020fairness} identified unfairness toward inactive users and proposed reranking approaches. We address segment bias within multi-objective optimization through segment-aware targeting that applies pre-calibrated weights per viewer lifecycle stage.

\section{Preliminaries}
\label{sec:preliminaries}

\subsection{Two-Stage Recommendation System}
\label{sec:twitch_recsys}
Twitch's recommendation system follows a two-stage Deep Neural Networks architecture \citep{covington2016deep}. The first stage employs a two-tower model to retrieve tens of channels from a catalog of millions of creators \citep{chen2022weighing}. The real-time ranking stage uses a deep neural network model to predict the minutes play probability given a viewer \(v_i\) and retrieved channels \(C_k, k=1, ..., n\). The system ranks the probabilities and shows the ordered channels to the viewers. In this work, we focus on the ranking stage.

\subsection{Multi-Objective Optimization For Twitch}
\label{sec:twitch_moo_problem}
While the base ranking model focused solely on engagement through predicting minutes play, Twitch's long-term business goals require optimizing multiple objectives simultaneously, including engagement, retention, and monetization. The real-time ranking model should optimize these high-level objectives by encouraging specific viewer actions, which serve as model training targets\footnote{Throughout this paper, we use \textit{objective} to refer to business goals, \textit{target} for viewer actions used in training, and \textit{task} for the corresponding prediction problems.}, and meanwhile stays adaptive to viewer segments. This section introduces Twitch's viewer actions and their correlation with the objectives, and defines viewer segments.

\paragraph{Viewer Actions and Objectives.}
Viewers on Twitch engage through five key actions: 
\begin{itemize}
\item Short-form Minutes Play (SMP): Playing a channel for $\tau_s$ minutes. SMP represents immediate viewer interest after clicking into a channel.
\item Long-form Minutes Play (LMP): Playing a channel for $\tau_l$ minutes, indicating deeper content engagement beyond initial curiosity.\footnote{$\tau_l \gg \tau_s$, where $\tau_l$ and $\tau_s$ are watch-time thresholds; exact values are proprietary.}
\item Chatting: in a channel's community.
\item Following: a channel for easier access and notifications.
\item Spending: on subscriptions and gifts to support a channel.
\end{itemize}
These actions serve different business objectives: \textit{SMP}, \textit{LMP}, \textit{chat}, and \textit{follow} primarily drive new viewer engagement and retention. Sustained engagement can indirectly lead to monetization through purchases and subscriptions. Direct \textit{spend} actions increase product revenue and indicate deeper viewer-channel bonds. This natural progression of viewer actions provides appropriate targets for optimizing engagement, retention, and revenue simultaneously.

\paragraph{Viewer Segments.}
Newer and established viewers behave differently. To better understand viewer behavior, Twitch segments viewers into two main categories based on engagement levels:

\textit{Early (E):} New (account age <= $M$ days) or less frequent (in the last $M$ days, chat or visit days <= $N$ days) viewers.\footnote{$M$ and $N$ are service-specific thresholds; exact values are proprietary.}

\textit{Dedicated (D):} Frequently engaging viewers and enthusiasts (account age > $M$ days, and in the last $M$ days, visit days > $N$).

\section{Methods}
\label{sec:methods}

We present our approach through progressive enhancements to address sparse delayed feedback and viewer segment bias. We begin with problem formulation (Section \ref{sec:methods_problem_formulation}), then introduce the delayed window (Section \ref{sec:methods_delayed_window_framework}), Multi-Model (Section \ref{sec:methods_multi_model_architecture}), Viewer Segment Targeting (Section \ref{sec:methods_viewer_segment_targeting}), and MMoE Enhancement (Section \ref{sec:methods_mmoe_enhancement}) with design rationale. Each stage was validated through comprehensive offline and online experiments.

\subsection{Problem Formulation}
\label{sec:methods_problem_formulation}
 Given a set of viewers $\mathcal{V}$ and channels $\mathcal{C}$, let $\mathbf{x}_{v_i,c_j}$ denote the feature vector that combines viewer and channel signals for viewer ${v_i} \in \mathcal{V}$ and channel ${c_j} \in \mathcal{C}$. To optimize multiple business objectives simultaneously, we construct a ranking function that combines predictions for all viewer actions $a \in \mathcal{A} = \{SMP, LMP, chat, follow, spend\}$:

\begin{equation}
\label{equation:ranking_function}
\begin{split}
F^{Ranking}(\mathbf{x}_{v_i,c_j}) &= \sum_{a \in \mathcal{A}}
w_{a} \cdot p_{a}(\mathbf{x}_{v_i,c_j})
\end{split}
\end{equation}

where $p_a(\mathbf{x}_{v_i,c_j})$ represents the predicted probability of viewer action $a$, and $w_a$ are the weights that balance different business objectives. While this formulation is straightforward, its implementation faces key challenges previously discussed: sparse and delayed feedback signals, and the need for segment-specific optimization. Additionally, scalability is critical -- Twitch serves millions of users daily, requiring efficient real-time inference. In the following sections, we present our solutions to address these challenges while maintaining real-time performance requirements.

\subsection{Delayed Window Framework}
\label{sec:methods_delayed_window_framework}
To address target sparsity and delayed feedback, we expand the observation window for positive signal collection. Aggregating across multiple days makes the sparse targets denser and captures delayed feedback. For sparse actions \textit{chat}, \textit{follow}, and \textit{spend}, we formalize this approach as follows. For a recommendation session at time $t$ where viewer $v_i$ is exposed to channel $c_j$, we define the delayed window $DW$ as:

\begin{equation}
\begin{split}
    DW_{v_i,c_j}(t, \Delta t) = \{(a, t') : a \in \mathcal{A}, t \leq t' \leq t + \Delta t,  \\ \text{viewer } v_i \text{ performs action } a \text{ on channel } c_j \text{ at time } t'\}
\end{split}
\end{equation}

where $t'$ represents any timestamp within the delayed window interval. For each action, this transforms sparse immediate observations into aggregated denser delayed engagement signals through:
\begin{equation}
\label{equation:delayed_feedback}
y_{v_i,c_j}^{delayed} = \mathbb{I}[|DW_{v_i,c_j}(t, \Delta t)| > 0]
\end{equation}
where $\mathbb{I}[\cdot]$ is the indicator function that equals 1 if the corresponding target action occurs within the delayed window, and 0 otherwise.

\subsection{Multi-Model Architecture}
\label{sec:methods_multi_model_architecture}

\begin{figure}
    \centering
    \includegraphics[width=1.0\linewidth]{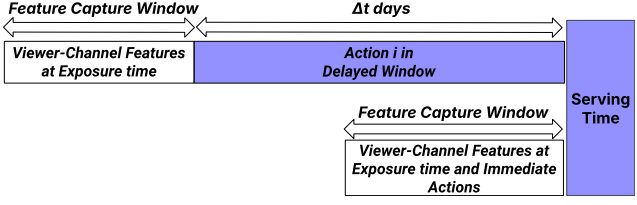}
    \caption{Timeline illustration of Delayed Window approach. Top: DSM training data uses exposure-time features with delayed window targets collected over $\Delta t$ days. Bottom: FSM training data uses exposure-time features and immediate actions, ensuring fresh information at serving.}
    \Description{Diagram showing feature capture window at exposure time and delayed action window spanning delta-t days for training label collection.}
    \label{fig:delayed_window_timeline}
\end{figure}

The $\Delta t$ delayed window effectively alleviates viewer action sparsity. However, relying solely on delayed signals presents a trade-off: while delayed feedback provides comprehensive action information, it may miss recent changes in viewer preferences that are captured by immediate signals at inference time. As illustrated in Figure \ref{fig:delayed_window_timeline}, using only delayed targets means that features would be $\Delta t$ days old by serving time, potentially missing recent viewer behavior patterns. To balance comprehensive historical feedback with fresh viewer preferences, we propose a multi-model (MM) architecture.

For less sparse viewer actions such as SMP, we use fresh signals without delayed windows and train Fresh Signal Models (FSM) to capture immediate engagement. The delayed window approach is only applied to sparse actions $a \in \{chat, follow, spend\}$, which are learned by Delayed Signal Models (DSM). Each FSM learns from dense and immediate actions, while each DSM is trained on delayed targets. All models are trained in parallel. During inference, both types of models output predictions $p_{a}(\mathbf{x}_{v_i,c_j})$ for their corresponding actions. This architecture provides the foundation for further enhancements in viewer segment targeting and model efficiency.

\subsection{Viewer Segment Targeting (VST)}
\label{sec:methods_viewer_segment_targeting}

With DSMs and FSMs providing action-specific predictions, we address viewer segment bias through segment-aware weighting of model outputs. Since D viewers contribute roughly 4x more training samples than E viewers in our data, model predictions are inherently skewed toward D viewer behavior. Using the same ensemble weights across all viewer segments would therefore propagate this bias into the final rankings. Moreover, business objectives vary across viewer lifecycle stages: engagement and retention are crucial for E viewers, while D viewers can be optimized for monetization, as monetization indicates deeper viewer-channel bonds and revenue. To address these varying optimization needs while maintaining operational simplicity, we apply lightweight segment-conditioned objective scalarization at inference time rather than training separate segment-specific ranking models.

\paragraph{Segment-Aware Ensemble.}
For each viewer $v_i$, we determine their segment $s \in \mathcal{S} = \{E, D\}$ based on account age and engagement level described in Section \ref{sec:twitch_moo_problem}. We then apply the segment-conditioned weights to the final ranking score:

\begin{equation}
\label{equation:final_vs_ranking_function}
\begin{split}
F^{VST}(\mathbf{x}_{v_i,c_j}) &= \sum_{a \in \mathcal{A}}
w_{a, s} \cdot p_{a}(\mathbf{x}_{v_i,c_j})
\end{split}
\end{equation}
where $w_{a, s}$ represent weights for action $a \in \mathcal{A}$ and viewer segment $s \in \mathcal{S}$. The models are trained on all the viewers, while during inference, we serve the predictions differently. This post-training weighting approach preserves deployment simplicity and online iteration stability while enabling segment-specific optimization.

\subsection{Multi-Gate Mixture-of-Experts Enhancement}
\label{sec:methods_mmoe_enhancement}
Building upon our multi-model architecture with viewer segment targeting, we further optimize the system's efficiency while maintaining its effectiveness. While the parallel models (Section \ref{sec:methods_multi_model_architecture}) effectively capture distinct signals for each target, maintaining separate models for each action increases system complexity and training costs. To address this, we consolidate multiple independent DSMs into a single multi-task learning model while preserving the benefits of segment-aware targeting. We experiment with Multi-gate Mixture-of-Experts (MMoE) \citep{ma2018modeling}, Shared-Bottom, and CGC \citep{tang2020progressive} (the core extraction module of PLE) as candidate MTL backbones in Section \ref{sec:experiments}; here we describe the architecture using MMoE as the representative example.

\begin{figure*}
    \centering
    \includegraphics[width=0.8\linewidth]{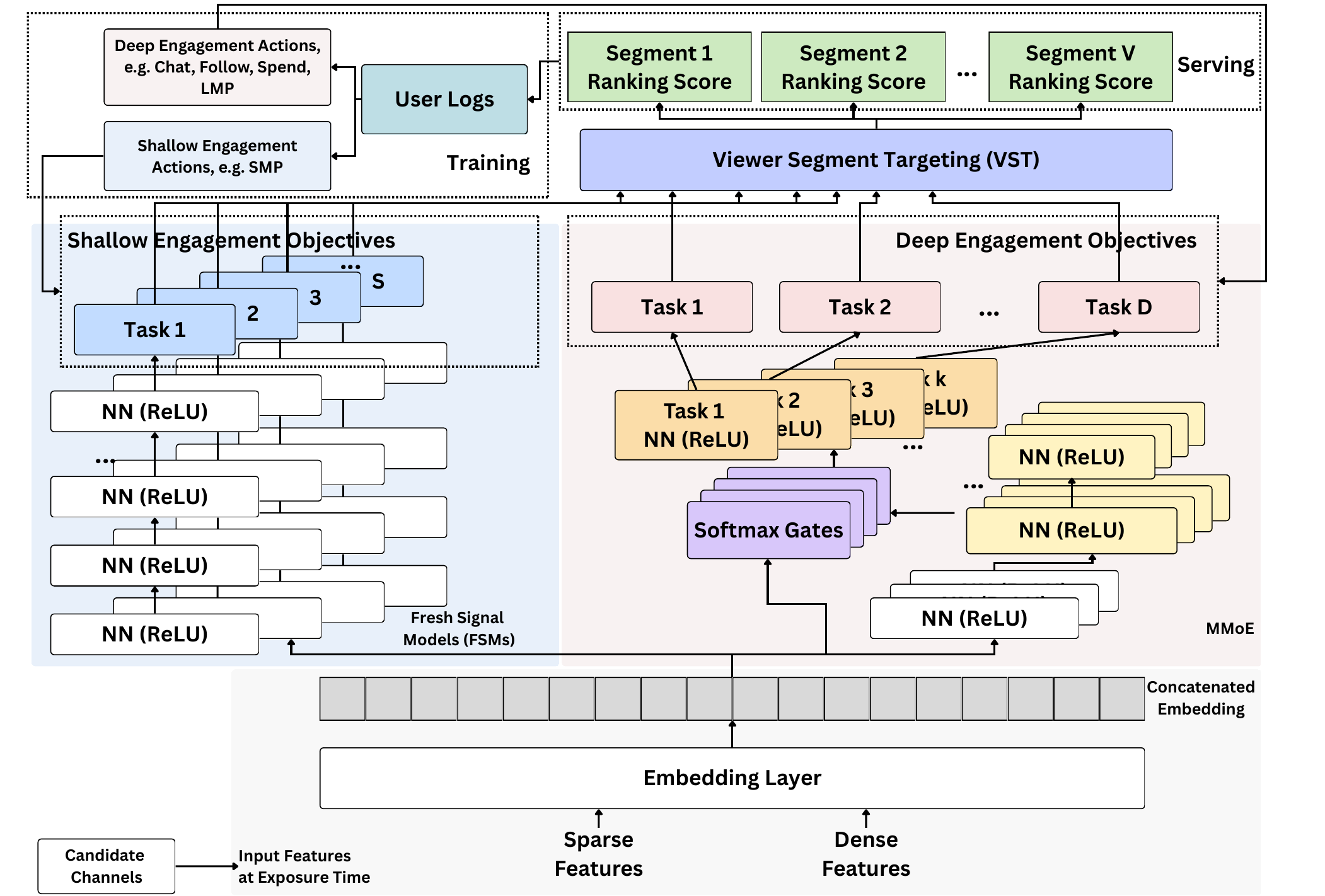}

    \caption{The FSM-MMoE-VST architecture. FSM handles shallow engagement (SMP); MMoE jointly models deep engagement tasks including immediate (LMP) and delayed targets (chat, follow, spend) with 14-day window aggregation. VST applies segment-conditioned weights at inference for Early (E) and Dedicated (D) viewers.}
    \Description{System architecture diagram showing two main components: left side has stacked Fresh Signal Models for shallow engagement with fully connected layers; right side shows MMoE with shared embedding layer, sparse and dense features, expert networks, softmax gates, and task-specific towers for deep engagement targets, connected to Viewer Segment Targeting module that produces segment-specific ranking scores.}
    \label{fig:proposed_architecture}
\end{figure*}

As illustrated in Figure \ref{fig:proposed_architecture}, our final FSM-MMoE-VST architecture combines a Fresh Signal Model for shallow engagement (SMP) with MMoE for joint modeling of deeper engagement, retention, and monetization targets (LMP, chat, follow, spend), integrated with the VST module at inference. Our task grouping is motivated by distinct characteristics of viewer actions:

\begin{itemize}
\item SMP (shallow engagement): Serves as a critical early indicator requiring immediate prediction, maintained as an independent FSM.
\item LMP (deep engagement): Represents sustained viewer interest beyond initial exploration. While LMP uses immediate labels rather than delayed windows, its role as a deeper engagement signal aligns naturally with delayed action targets, making it suitable for joint modeling.
\item Delayed actions (chat, follow, spend): Represent community engagement, retention, and monetization objectives, captured using 14-day delayed windows to address sparsity.
\end{itemize}

Based on these characteristics, our final architecture combines an independent FSM for SMP with an MMoE model ($K=4$ expert networks) that jointly handles $\{LMP, chat, follow, spend\}$.\footnote{Number of experts selected via hyperparameter optimization (HPO).} MMoE's task-specific gating mechanisms allow different targets to selectively leverage shared expert representations while maintaining specialization. The segment-aware targeting is preserved through segment-specific weighting at inference.

Formally, our final ranking function combines predictions from both FSM and MMoE:

\begin{equation}
\begin{split}
F^{FSM-MMOE-VST}(\mathbf{x}_{v_i,c_j}, s) = w_{smp,s} \cdot p_{smp}^{FSM}(\mathbf{x}_{v_i,c_j}) + \\ 
\sum_{a} w_{a,s} \cdot p_a^{MMoE}(\mathbf{x}_{v_i,c_j})
\end{split}
\end{equation}

where $a \in \{LMP,chat,follow,spend\}$, $s \in \mathcal{S} = \{E, D\}$ is the viewer segment, and $w_{smp,s}$ and $w_{a,s}$ are segment-specific weights for FSM and MMoE outputs respectively.

\section{Experiments}
\label{sec:experiments}
\subsection{Experimental Setup}
\subsubsection{Delayed Window Analysis.}
\label{sec:window_analysis}
To determine the optimal delayed window size for sparse targets (chat, follow, spend), we analyzed positive sample ratios across windows of \{7, 14, 21, 28, 35\} days using historical impressions from 6 million viewers in the training dataset. Weekly intervals were chosen to balance captured signals with data aggregation efficiency because Twitch users present different behaviors in weekdays and weekends. We observed substantial density improvements with the delayed window approach: chat targets increase by around 9 times in positive labels with just a 7-day window and by 12 times at 14 days. However, extending windows indefinitely introduces diminishing returns and raises concerns about data aggregation computation and signal staleness. In our final implementation, we apply the delayed window to the sparse targets \textit{chat}, \textit{follow}, and \textit{spend}, while \textit{SMP} and \textit{LMP} use immediate labels.

\subsubsection{Model Architecture and Training.}
\label{sec:model_arc_details}

We used 7 days of historical impression data from 6 million viewers for training. For the Fresh Signal Model (FSM), \textit{SMP} is trained on all impressions with immediate targets. For the MMoE component, \textit{LMP} uses immediate labels, while delayed targets (\textit{chat}, \textit{follow}, \textit{spend}) use a 14-day delayed window. Our evaluation dataset comprises 1-day historical data from 1 million viewers, with a 35-day forward window for ground truth labels to ensure comprehensive coverage of delayed actions\footnote{Spending actions, such as subscriptions, can be monthly.}.

In the multi-model architecture, the FSM uses four fully-connected layers (size=1000) while DSMs use four layers (size=300), all with ReLU activation. For the MMoE-enhanced architecture, we maintain the same FSM and replace the independent DSMs with an MMoE model consisting of three shared bottom layers (size=1000), four expert networks (two layers: 512, 256), task-specific gating networks (four layers: 128 each), and one-layer task towers (size=64). We also evaluate Shared-Bottom and CGC \citep{tang2020progressive} as alternative MTL backbones, in both single unified (all five targets) and FSM-based (paired with an independent SMP model) configurations. We exclude sequential conversion models such as ESMM \citep{ma2018entire} and AITM \citep{xi2021modeling} as they assume sequential action funnels (e.g., click $\rightarrow$ purchase), whereas live-streaming viewers' important actions occur independently without a fixed order. All models use binary cross entropy loss and process the same feature set, including viewer demographics, channel characteristics, and viewer-channel interactions. We use the Adam optimizer with learning rates in [0.001, 0.005, 0.01] and batch sizes in [1024, 2048], selected through hyperparameter optimization. We use NDCG on the validation set for model selection.

\subsubsection{Viewer Segment Targeting Strategy.}
\label{sec:viewer_segment_module_enhancement}
  We hypothesize that E and D viewers have different optimization needs. E viewers, being newer and less familiar with streamers and their communities, benefit most from recommendations focused on immediate engagement (\textit{SMP}). In contrast, D viewers can be optimized for deeper engagement (\textit{chat}, \textit{follow}, \textit{LMP}) as well as monetization (\textit{spend}).

  To counteract training data bias where E viewers contribute fewer samples, we considered applying higher loss weights to E viewers' samples during training. However, this approach did not improve performance and introduced additional training complexity in our context. Instead, we apply differentiated weighting strategies at inference time. For E viewers, we prioritize \textit{SMP} with minimal weights on other actions. For D viewers, we balance multiple objectives. We selected candidate weights in offline tests and finalized the production configuration through iterative A/B testing on business metrics and guardrails. Specific weight values are proprietary but follow the prioritization strategy described above. This approach allows us to tailor recommendations to viewer lifecycle stages -- in the early stage, we encourage them to watch and connect with the streamers they are interested in; later on, we encourage them to build deeper bonds by interacting with streamers. Importantly, this segment-aware behavior is achieved purely through differentiated weighting at inference, avoiding the complexity of maintaining separate segment-specific models.

\begin{figure*}
    \centering
    \includegraphics[width=0.9\linewidth]{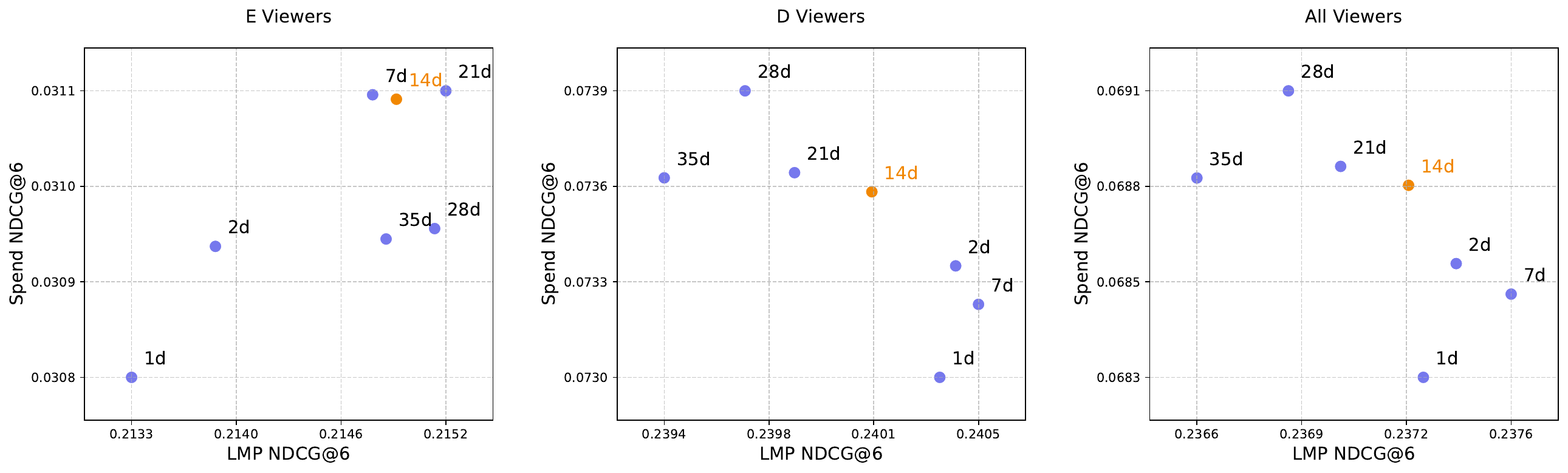}
    \caption{Comparison of LMP and Spend NDCG@6 across delayed window sizes (1-35 days), using consistent multi-model architecture and features. 14-day window highlighted.}
    \Description{Three scatter plots for E Viewers, D Viewers, and All Viewers showing LMP NDCG at 6 on x-axis versus Spend NDCG at 6 on y-axis, with points labeled by delayed window size from 1 to 35 days, with 14-day window highlighted.}
    \label{fig:window_selection_scatter}
\end{figure*}

\subsection{Offline Results}
\subsubsection{Delayed Window Selection}
Using a 2025 June-July dataset, we analyzed the impact of different delayed window sizes on the multi-model performance. Figure \ref{fig:window_selection_scatter} shows NDCG@6 results for both engagement (measured by LMP - Long Minutes Play) and monetization (measured by Spend) across different window lengths.\footnote{Twitch's left navigation displays 6 recommended channels by default} The analysis reveals distinct patterns across viewer segments:

For E viewers, the 1-day window capturing immediate feedback performed worst, while extending to 7-14-21 days significantly improved both LMP and Spend NDCG@6. This suggests that aggregating delayed feedback helps better understand newer viewers' behavior patterns. 

For D viewers, we observe an interesting trade-off: shorter windows (1-7 days) excel in LMP NDCG@6 but underperform in Spend NDCG@6 compared to longer windows. This pattern indicates that while longer aggregation periods help capture sparse, delayed actions like spending, excessive window lengths introduce noise by attributing actions to impressions they are no longer associated with, degrading engagement signal quality.

Based on these results, we identify 14 days as the optimal window length, achieving significant spend improvements while maintaining strong engagement performance. Using a 21-day window leads to similar performance as 14 days, with slightly better Spend and slightly worse LMP NDCG@6. However, considering the data aggregation computation and staleness, we still choose 14 days. This analysis demonstrates that appropriate delayed feedback aggregation can enhance multi-objective learning effectiveness, particularly for capturing sparse actions without compromising immediate engagement signals.

\begin{table*}[h]
    \centering
    \caption{Offline NDCG@6 comparing different architectures. All models use consistent feature sets and 14-day delayed window where applicable. E and D represent early and dedicated viewers respectively. Bold indicates best performance per column.}
    \begin{tabular}{lcccccc}
    \toprule
    & \multicolumn{3}{c}{\textbf{LMP}} & \multicolumn{3}{c}{\textbf{Spend}} \\
    \cmidrule(r){2-4} \cmidrule(l){5-7}
    \textbf{Model} & \textbf{E} & \textbf{D} & \textbf{All} & \textbf{E} & \textbf{D} & \textbf{All} \\
    \midrule
    YouTube DNN \citep{covington2016deep} & 0.2090 & 0.2504 & 0.2459 & 0.0282 & 0.0780 & 0.0726 \\
    \midrule
    MM + Delayed & 0.2071 & 0.2495 & 0.2448 & 0.0283 & \textbf{0.0797} & \textbf{0.0741} \\
    MM + Delayed + VST & 0.2090 & 0.2498 & 0.2453 & 0.0283 & 0.0795 & 0.0740 \\
    \midrule
    Single Shared Bottom & 0.1967 & 0.2402 & 0.2355 & 0.0275 & 0.0794 & 0.0738 \\
    Single MMoE & 0.1961 & 0.2395 & 0.2347 & 0.0276 & \textbf{0.0797} & \textbf{0.0741} \\
    Single CGC \citep{tang2020progressive} & 0.1964 & 0.2403 & 0.2355 & 0.0277 & 0.0795 & 0.0739 \\
    \midrule
    FSM + Shared Bottom + VST & 0.2090 & \textbf{0.2509} & \textbf{0.2463} & 0.0283 & 0.0796 & 0.0740 \\
    FSM + MMoE + VST & \textbf{0.2095} & 0.2507 & 0.2462 & \textbf{0.0284} & 0.0794 & 0.0739 \\
    FSM + CGC + VST & \textbf{0.2095} & 0.2507 & 0.2462 & 0.0282 & 0.0794 & 0.0738 \\
    \bottomrule
    \end{tabular}
\label{tab:offline_model_comparison}
\end{table*}

\subsubsection{Model Performance}
With a 14-day delayed window, we experimented with our proposed architectures. The baseline model is a single-objective point-wise deep neural network modeling SMP trained on Twitch dataset. According to Table \ref{tab:offline_model_comparison}, our architectural progression shows clear improvements.

First, MM with delayed window improved overall Spend NDCG@6 by 2.07\% (0.0726 to 0.0741), primarily driven by D viewers (0.0780 to 0.0797), while incurring a slight LMP NDCG@6 decrease (-0.45\% overall). Adding VST successfully recovered the LMP performance, particularly for E viewers, while maintaining the Spend improvements. This validates our strategy of prioritizing engagement for growth-potential E viewers while preserving monetization for established D viewers.

Comparing single unified MTL models (rows 4-6) against the DNN baseline reveals a consistent pattern: jointly modeling all five targets in a single architecture severely degrades LMP performance (-4.2\% to -4.6\%) regardless of the MTL backbone (MMoE, Shared-Bottom, or CGC), demonstrating that shallow engagement (SMP) conflicts with deeper engagement and delayed targets when modeled together.

Among FSM-MTL based architectures (rows 7-9), all three MTL backbones achieve comparable offline performance, suggesting that the fresh/delayed signal separation is the primary driver of improvement rather than the specific multi-task architecture. We evaluated both Shared-Bottom and MMoE in online A/B testing, where MMoE demonstrated stronger performance (Section \ref{sec:experiments_online}); CGC, which showed comparable offline results, remains a candidate for future online evaluation.

All three FSM-MTL based architectures improve over the DNN baseline, with FSM + MMoE + VST improving LMP NDCG@6 by 0.12\% (0.2459 to 0.2462) and Spend NDCG@6 by 1.79\% (0.0726 to 0.0739), with gains across both viewer segments (E LMP: +0.24\%, D LMP: +0.12\%). The comparable performance across MTL backbones reinforces our key finding: the architectural separation of fresh and delayed signals is the dominant factor in our context.

\subsection{Online Experiments}
\label{sec:experiments_online}

Based on the promising offline results, we conducted three 14-day A/B experiments to evaluate our approach stage by stage.\footnote{Follow metrics are omitted from Exp.~1-2 for brevity.}

\begin{table}[h]
\centering
\caption{Online A/B experiment results across three stages. All metrics are CUPED \citep{deng2013improving}. *$p < 0.05$, **$p < 0.01$.}
\begin{tabular}{clcccc}
\toprule
\textbf{Exp.} & \textbf{Segment} & \textbf{ARPU} & \textbf{DAV} & \textbf{LMP} & \textbf{Follow} \\
\midrule
1 & E & $-$0.72\% & $+$0.10\%** & $+$0.11\% & -- \\
1 & D & $+$0.56\%* & $+$0.08\%** & $+$0.17\%** & -- \\
1 & All & $+$0.34\% & $+$0.09\%** & $+$0.16\%** & -- \\
\midrule
2 & E & $-$0.66\% & $+$0.15\%* & $+$0.25\%** & -- \\
2 & D & $+$0.33\% & $-$0.01\% & $-$0.09\% & -- \\
2 & All & $+$0.23\% & $+$0.03\% & $+$0.07\% & -- \\
\midrule
3 & E & $+$1.45\% & $+$0.09\% & $+$0.12\% & $+$0.16\% \\
3 & D & $-$0.35\% & $+$0.06\% & $+$0.05\% & $+$0.44\%** \\
3 & All & $-$0.01\% & $+$0.08\%* & $+$0.10\%* & $+$0.27\%** \\
\bottomrule
\end{tabular}
\label{tab:online_results}
\end{table}

\begin{table*}[h]
    \centering
    \caption{Ablation study results comparing MMoE task grouping strategies. All the delayed targets use a 14-day window. Single MMoE has no viewer segment targeting; all other variants include VST. Bold indicates best performance for each metric.}
    \begin{tabular}{lccccccc}
    \toprule
    & \multicolumn{3}{c}{\textbf{LMP}} & \multicolumn{3}{c}{\textbf{Spend}} \\
    \cmidrule(r){2-4} \cmidrule(l){5-7}
    \textbf{Architecture} & \textbf{E} & \textbf{D} & \textbf{All} & \textbf{E} & \textbf{D} & \textbf{All} \\
    \midrule
    Single MMoE (All targets) & 0.1961 & 0.2395 & 0.2347 & 0.0276 & \textbf{0.0797} & \textbf{0.0741} \\
    FSM(SMP) + MMoE(Delayed only) & 0.2091 & 0.2501 & 0.2456 & 0.0280 & 0.0791 & 0.0736 \\
    MMoE(SMP+LMP) + MMoE(Delayed) & 0.2094 & 0.2505 & 0.2460 & 0.0282 & 0.0794 & 0.0739 \\
    FSM(SMP) + MMoE(LMP+Delayed) & \textbf{0.2095} & \textbf{0.2507} & \textbf{0.2462} & \textbf{0.0284} & 0.0794 & 0.0739 \\
    \bottomrule
    \end{tabular}
\label{tab:ablation_results}
\end{table*}
  
\paragraph{Experiment 1: Multi-Model with Delayed Targets.}
\label{sec:experiment_1}

The baseline uses a single-objective deep neural network optimizing SMP with a monetization heuristic feature, Monetization Attach Ratio (MAR), which was the previous A/B test winner for revenue optimization. Our treatment implements the multi-model architecture: an FSM for immediate engagement (SMP) combined with independent DSMs for delayed targets (chat, follow, spend) using a 14-day delayed window. The experiment impacted millions of visitors, with results in Table \ref{tab:online_results} (Exp.\ 1). We capped ARPU at a fixed daily threshold to reduce variance from high spenders.

The results demonstrate that our multi-model with delayed targets significantly improves metrics across segments: overall DAV (+0.09\%, $p < 0.01$) and LMP (+0.16\%, $p < 0.01$), translating to millions of annual active viewer days. D viewers show both monetization (+0.56\% ARPU, $p < 0.05$) and engagement gains, while E viewers improve in engagement (+0.10\% DAV, $p < 0.01$) but directionally decline in monetization (-0.72\% ARPU). Since D viewers contribute the majority of commerce revenue, the +0.56\% capped ARPU improvement represents significant business impact. Notably, this ARPU improvement is achieved over a baseline whose monetization heuristic (MAR) was the previous A/B test winner, making the gains attributable to the multi-model architecture rather than removal of a weak heuristic.

\paragraph{Experiment 2: Viewer Segment Targeting.}
Using Experiment 1's winner as baseline, we tested segment-aware targeting (Equation \ref{equation:final_vs_ranking_function}) where E and D viewers receive segment-conditioned weights.

Our segment-aware approach successfully improves E viewer engagement (+0.15\% DAV, +0.25\% LMP) while maintaining D performance in the A/B (Table \ref{tab:online_results}, Exp.\ 2) with the best-performing weights. While overall metrics show modest gains (+0.03\% DAV, +0.07\% LMP) due to D viewers' slight changes, the improved E engagement represents valuable long-term growth potential.

\paragraph{Experiment 3: MMoE Enhancement.} We then used the viewer segment targeting variant as the baseline, and enhanced our architecture by replacing multiple independent models with an FSM + MMoE + VST combination that jointly handles multiple targets.

The MMoE enhancement in Table \ref{tab:online_results} (Exp.\ 3) shows significant improvements across multiple metrics: overall total follows (+0.27\%, $p < 0.01$), overall DAV (+0.08\%, $p < 0.05$), and overall LMP (+0.10\%, $p < 0.05$). The D DAV improvement (+0.06\%, $p = 0.11$), while not meeting the significance threshold, demonstrates a positive trend. We also evaluated FSM + Shared-Bottom + VST in the same experiment, which showed weaker results: overall capped ARPU +0.19\%, DAV +0.06\% ($p = 0.10$), LMP +0.06\%, and follows +0.15\%, none reaching statistical significance, confirming MMoE as the stronger MTL backbone in our online setting.

Meanwhile, the MMoE consolidation reduces the delayed-target modeling component from 26.7 million parameters (multiple independent DSMs) to approximately 15.5 million parameters (an MMoE), a 41.9\% reduction that significantly decreases training costs while maintaining serving latency. This demonstrates significant efficiency gains at scale for a real-time inference ranking model.

\section{Ablation Studies}
We conduct ablation studies to validate our architectural design decisions for the MMoE component. While Table \ref{tab:offline_model_comparison} compares across architecture families, Table \ref{tab:ablation_results} focuses on how targets should be grouped when using the FSM + MMoE design. 

The Single MMoE approach that jointly models all five targets shows a significant trade-off: while achieving the highest overall spend NDCG@6 (0.0741), it severely degrades LMP performance (-4.7\% from 0.2462 to 0.2347). This confirms our hypothesis that including SMP in joint modeling hurts engagement metrics due to its dominant signal overwhelming sparser targets. 

Among FSM-based architectures, we compare three strategies. Excluding LMP from MMoE (row 2) achieves moderate performance. The dual-MMoE approach (row 3) that groups SMP and LMP in one MMoE while keeping delayed targets in another shows slight LMP improvement but increased system complexity with minimal gains. Our final architecture (row 4) that includes LMP with delayed targets in MMoE achieves the best LMP performance (0.2462) while maintaining competitive spend NDCG (0.0739). This supports our hypothesis that LMP, despite being more frequent, shares underlying engagement patterns with delayed actions that benefit from joint modeling.

\section{Discussion}
\label{sec:discussion}

\subsection{Signal Separation vs. Architecture Choice}
A recurring finding across our experiments is that the architectural separation of fresh and delayed signals matters more than the choice of MTL backbone. All three MTL architectures (MMoE, Shared-Bottom, CGC) achieved comparable offline performance when paired with an independent FSM, while all single unified models severely degraded engagement regardless of backbone sophistication. We attribute this to the fundamental differences between shallow engagement (SMP) and deeper engagement and monetization targets: they differ in both engagement depth and target density, with SMP being orders of magnitude denser than actions like follow and spend. When modeled jointly, the dense shallow signal dominates gradient updates and overwhelms sparser targets. This suggests that when targets differ substantially in density and engagement depth, practitioners can prioritize signal-level architectural separation over MTL backbone selection.

\subsection{Practical Design Lessons}
We initially explored training-time interventions to address viewer segment bias, including up-weighting E samples in the loss function. This did not improve performance. In contrast, inference-time segment weighting with zero additional parameters proved effective for E engagement without degrading D performance. This framework naturally extends to finer-grained segmentation, though automating weight configuration through online learning remains an open challenge. For the delayed window, extending the delayed window improves target density up to a point but introduces trade-offs: longer windows increase data aggregation costs and risk capturing stale patterns. Our analysis showed diminishing returns beyond 14 days, with D engagement metrics declining at longer windows. Additionally, longer windows increase pipeline dependency, as the training data cannot be finalized until $\Delta t$ days after exposure.

\section{Conclusion and Future Work}
We present a multi-objective optimization framework for live-streaming recommendations that addresses target sparsity, delayed feedback, and viewer segment bias through: (1) delayed window aggregation for sparse targets, (2) multi-model architecture balancing fresh and delayed signals, (3) segment-aware targeting across viewer lifecycle stages, and (4) MMoE integration that improves performance while reducing parameters. Through staged experiments on Twitch, we achieved +0.09\% Daily Active Viewers and +0.56\% D ARPU with the delayed window and multi-model approach, with segment targeting further improving E DAV by +0.15\% and MMoE adding +0.08\% overall DAV, achieving sub-110ms p99 ranking latency at scale. The approach also generalized to Twitch's mobile livefeed, achieving +1.12\% positive interactions in a 14-day A/B test ($p$ < 0.001). Future directions include automating weight optimization through online learning, exploring fine-grained segmentation beyond E/D categories, and investigating adaptive delayed windows.

\begin{acks}
We gratefully acknowledge Edgar Chen, whose early work on the delayed window concept, initial experimentation, and data curation formed a cornerstone of the multi-objective framework presented in this paper. His guidance on the system architecture, as well as his leadership in the early MMoE exploration through mentorship, were instrumental to this work. We also thank Chad Mills for initiating the research direction that led to this work, for his contributions through insightful discussions and input throughout the project, and for his dedicated review of this paper.
\end{acks}

\bibliographystyle{ACM-Reference-Format}
\bibliography{bibfile}

@inproceedings{zhao2019recommending,
  title={Recommending what video to watch next: a multitask ranking system},
  author={Zhao, Zhe and Hong, Lichan and Wei, Li and Chen, Jilin and Nath, Aniruddh and Andrews, Shawn and Kumthekar, Aditee and Sathiamoorthy, Maheswaran and Yi, Xinyang and Chi, Ed},
  booktitle={Proceedings of the 13th ACM conference on recommender systems},
  pages={43--51},
  year={2019}
}

@inproceedings{xie2021personalized,
  title={Personalized approximate pareto-efficient recommendation},
  author={Xie, Ruobing and Liu, Yanlei and Zhang, Shaoliang and Wang, Rui and Xia, Feng and Lin, Leyu},
  booktitle={Proceedings of the web conference 2021},
  pages={3839--3849},
  year={2021}
}

@inproceedings{lin2019pareto,
  title={A pareto-efficient algorithm for multiple objective optimization in e-commerce recommendation},
  author={Lin, Xiao and Chen, Hongjie and Pei, Changhua and Sun, Fei and Xiao, Xuanji and Sun, Hanxiao and Zhang, Yongfeng and Ou, Wenwu and Jiang, Peng},
  booktitle={Proceedings of the 13th ACM Conference on recommender systems},
  pages={20--28},
  year={2019}
}

@inproceedings{ma2018modeling,
  title={Modeling task relationships in multi-task learning with multi-gate mixture-of-experts},
  author={Ma, Jiaqi and Zhao, Zhe and Yi, Xinyang and Chen, Jilin and Hong, Lichan and Chi, Ed H},
  booktitle={Proceedings of the 24th ACM SIGKDD international conference on knowledge discovery \& data mining},
  pages={1930--1939},
  year={2018}
}

@inproceedings{bai2024gradcraft,
  title={GradCraft: Elevating Multi-task Recommendations through Holistic Gradient Crafting},
  author={Bai, Yimeng and Zhang, Yang and Feng, Fuli and Lu, Jing and Zang, Xiaoxue and Lei, Chenyi and Song, Yang},
  booktitle={Proceedings of the 30th ACM SIGKDD Conference on Knowledge Discovery and Data Mining},
  pages={4774--4783},
  year={2024}
}

@article{ruder2017overview,
  title={An overview of multi-task learning in deep neural networks},
  author={Ruder, Sebastian},
  journal={arXiv preprint arXiv:1706.05098},
  year={2017}
}

@inproceedings{ma2019snr,
  title={Snr: Sub-network routing for flexible parameter sharing in multi-task learning},
  author={Ma, Jiaqi and Zhao, Zhe and Chen, Jilin and Li, Ang and Hong, Lichan and Chi, Ed H},
  booktitle={Proceedings of the AAAI conference on artificial intelligence},
  volume={33},
  number={01},
  pages={216--223},
  year={2019}
}

@inproceedings{tang2020progressive,
  title={Progressive layered extraction (ple): A novel multi-task learning (mtl) model for personalized recommendations},
  author={Tang, Hongyan and Liu, Junning and Zhao, Ming and Gong, Xudong},
  booktitle={Proceedings of the 14th ACM conference on recommender systems},
  pages={269--278},
  year={2020}
}

@inproceedings{ding2021mssm,
  title={MSSM: a multiple-level sparse sharing model for efficient multi-task learning},
  author={Ding, Ke and Dong, Xin and He, Yong and Cheng, Lei and Fu, Chilin and Huan, Zhaoxin and Li, Hai and Yan, Tan and Zhang, Liang and Zhang, Xiaolu and others},
  booktitle={Proceedings of the 44th International ACM SIGIR Conference on Research and Development in Information Retrieval},
  pages={2237--2241},
  year={2021}
}

@inproceedings{wang2025home,
  title={Home: Hierarchy of multi-gate experts for multi-task learning at kuaishou},
  author={Wang, Xu and Cao, Jiangxia and Fu, Zhiyi and Gai, Kun and Zhou, Guorui},
  booktitle={Proceedings of the 31st ACM SIGKDD Conference on Knowledge Discovery and Data Mining V. 1},
  pages={2638--2647},
  year={2025}
}

@article{yu2020gradient,
  title={Gradient surgery for multi-task learning},
  author={Yu, Tianhe and Kumar, Saurabh and Gupta, Abhishek and Levine, Sergey and Hausman, Karol and Finn, Chelsea},
  journal={Advances in neural information processing systems},
  volume={33},
  pages={5824--5836},
  year={2020}
}

@inproceedings{chapelle2014modeling,
  title={Modeling delayed feedback in display advertising},
  author={Chapelle, Olivier},
  booktitle={Proceedings of the 20th ACM SIGKDD international conference on Knowledge discovery and data mining},
  pages={1097--1105},
  year={2014}
}

@inproceedings{ma2018entire,
  title={Entire space multi-task model: An effective approach for estimating post-click conversion rate},
  author={Ma, Xiao and Zhao, Liqin and Huang, Guan and Wang, Zhi and Hu, Zelin and Zhu, Xiaoqiang and Gai, Kun},
  booktitle={The 41st International ACM SIGIR Conference on Research \& Development in Information Retrieval},
  pages={1137--1140},
  year={2018}
}

@inproceedings{covington2016deep,
  title={Deep neural networks for youtube recommendations},
  author={Covington, Paul and Adams, Jay and Sargin, Emre},
  booktitle={Proceedings of the 10th ACM conference on recommender systems},
  pages={191--198},
  year={2016}
}

@inproceedings{chen2022weighing,
    title={Weighing dynamic availability and consumption for Twitch recommendations},
    author={Chen, Edgar and Ally, Mark and Santana, Eder and Ali, Saad},
    booktitle={KDD Workshop on Online and Adaptive Recommender Systems (OARS)},                   
    year={2022}     
}

@inproceedings{standley2020tasks,
  title={Which tasks should be learned together in multi-task learning?},
  author={Standley, Trevor and Zamir, Amir and Chen, Dawn and Guibas, Leonidas and Malik, Jitendra and Savarese, Silvio},
  booktitle={International conference on machine learning},
  pages={9120--9132},
  year={2020},
  organization={PMLR}
}

@inproceedings{jeunen2024multi,
  title={Multi-objective recommendation via multivariate policy learning},
  author={Jeunen, Olivier and Mandav, Jatin and Potapov, Ivan and Agarwal, Nakul and Vaid, Sourabh and Shi, Wenzhe and Ustimenko, Aleksei},
  booktitle={Proceedings of the 18th ACM Conference on Recommender Systems},
  pages={712--721},
  year={2024}
}

@inproceedings{ktena2019addressing,
  title={Addressing delayed feedback for continuous training with neural networks in CTR prediction},
  author={Ktena, Sofia Ira and Tejani, Alykhan and Theis, Lucas and Myana, Pranay Kumar and Dilipkumar, Deepak and Husz{\'a}r, Ferenc and Yoo, Steven and Shi, Wenzhe},
  booktitle={Proceedings of the 13th ACM conference on recommender systems},
  pages={187--195},
  year={2019}
}

@article{yoshikawa2018nonparametric,
  title={A nonparametric delayed feedback model for conversion rate prediction},
  author={Yoshikawa, Yuya and Imai, Yusaku},
  journal={arXiv preprint arXiv:1802.00255},
  year={2018}
}

@inproceedings{schnabel2016recommendations,
  title={Recommendations as treatments: Debiasing learning and evaluation},
  author={Schnabel, Tobias and Swaminathan, Adith and Singh, Ashudeep and Chandak, Navin and Joachims, Thorsten},
  booktitle={international conference on machine learning},
  pages={1670--1679},
  year={2016},
  organization={PMLR}
}

@article{chen2023bias,
  title={Bias and debias in recommender system: A survey and future directions},
  author={Chen, Jiawei and Dong, Hande and Wang, Xiang and Feng, Fuli and Wang, Meng and He, Xiangnan},
  journal={ACM Transactions on Information Systems},
  volume={41},
  number={3},
  pages={1--39},
  year={2023},
  publisher={ACM New York, NY}
}

@inproceedings{schnabel2020debiasing,
  title={Debiasing item-to-item recommendations with small annotated datasets},
  author={Schnabel, Tobias and Bennett, Paul N},
  booktitle={Proceedings of the 14th ACM Conference on Recommender Systems},
  pages={73--81},
  year={2020}
}

@inproceedings{ai2018unbiased,
  title={Unbiased learning to rank with unbiased propensity estimation},
  author={Ai, Qingyao and Bi, Keping and Luo, Cheng and Guo, Jiafeng and Croft, W Bruce},
  booktitle={The 41st international ACM SIGIR conference on research \& development in information retrieval},
  pages={385--394},
  year={2018}
}

@inproceedings{fu2020fairness,
  title={Fairness-aware explainable recommendation over knowledge graphs},
  author={Fu, Zuohui and Xian, Yikun and Gao, Ruoyuan and Zhao, Jieyu and Huang, Qiaoying and Ge, Yingqiang and Xu, Shuyuan and Geng, Shijie and Shah, Chirag and Zhang, Yongfeng and others},
  booktitle={Proceedings of the 43rd international ACM SIGIR conference on research and development in information retrieval},
  pages={69--78},
  year={2020}
}

@article{gao2022multi,
  title={Multi-Head Online Learning for Delayed Feedback Modeling},
  author={Gao, Hui and Yang, Yihan},
  journal={arXiv preprint arXiv:2205.12406},
  year={2022}
}

@inproceedings{agah2025pareto,
  title={Pareto-Optimal Solution: Optimizing Engagement and Revenue},
  author={Agah, Shaghayegh and Schaeffer, Shaun and Peifer, Maria and Sharma, Neeraj and Maheshwari, Ankit and Hamidian, Sardar},
  booktitle={Proceedings of the Nineteenth ACM Conference on Recommender Systems},
  pages={1034--1037},
  year={2025}
}

@inproceedings{shen2025paragon,
  title={Paragon: Parameter Generation for Controllable Multi-Task Recommendation},
  author={Shen, Chenglei and Zhao, Jiahao and Zhang, Xiao and Yu, Weijie and He, Ming and Fan, Jianping},
  booktitle={Proceedings of the Nineteenth ACM Conference on Recommender Systems},
  pages={370--380},
  year={2025}
}

@inproceedings{xi2021modeling,
  title={Modeling the sequential dependence among audience multi-step conversions with multi-task learning in targeted display advertising},
  author={Xi, Dongbo and Chen, Zhen and Yan, Peng and Zhang, Yinger and Zhu, Yongchun and Zhuang, Fuzhen and Chen, Yu},
  booktitle={Proceedings of the 27th ACM SIGKDD Conference on Knowledge Discovery \& Data Mining},
  pages={3745--3755},
  year={2021}
}

@inproceedings{nikolakaki2025semorec,
  title={SEMORec: A Scalarized Efficient Multi-Objective Recommendation Framework},
  author={Nikolakaki, Sofia Maria and Ma, Siyong and Chennu, Srivas and Topcu Altintas, Humeyra},
  booktitle={Proceedings of the Nineteenth ACM Conference on Recommender Systems},
  pages={1074--1077},
  year={2025}
}

@inproceedings{dishi2025practical,
  title={Practical Multi-Task Learning for Rare Conversions in Ad Tech},
  author={Dishi, Yuval and Friedler, Ophir and Karni, Yonatan and Silberstein, Natalia and Stolin, Yulia},
  booktitle={Proceedings of the Nineteenth ACM Conference on Recommender Systems},
  pages={1042--1045},
  year={2025}
}

@inproceedings{wang2023unbiased,
  title={Unbiased delayed feedback label correction for conversion rate prediction},
  author={Wang, Yifan and Sun, Peijie and Zhang, Min and Jia, Qinglin and Li, Jingjie and Ma, Shaoping},
  booktitle={Proceedings of the 29th ACM SIGKDD Conference on Knowledge Discovery and Data Mining},
  pages={2456--2466},
  year={2023}
}

@inproceedings{deng2013improving,
  title={Improving the sensitivity of online controlled experiments by utilizing pre-experiment data},
  author={Deng, Alex and Xu, Ya and Kohavi, Ron and Walker, Toby},
  booktitle={Proceedings of the sixth ACM international conference on Web search and data mining},
  pages={123--132},
  year={2013}
}

@inproceedings{liu2025multi,
  title={Multi-task Offline Reinforcement Learning for Online Advertising in Recommender Systems},
  author={Liu, Langming and Wang, Wanyu and Zhang, Chi and Li, Bo and Yin, Hongzhi and Wei, Xuetao and Su, Wenbo and Zheng, Bo and Zhao, Xiangyu},
  booktitle={Proceedings of the 31st ACM SIGKDD Conference on Knowledge Discovery and Data Mining V. 2},
  pages={4635--4646},
  year={2025}
}

@article{xia2026harmonrank,
  title={HarmonRank: Ranking-aligned Multi-objective Ensemble for Live-streaming E-commerce Recommendation},
  author={Xia, Boyang and Yu, Zhou and Zhu, Zhiliang and Sun, Hanxiao and Han, Biyun and Wang, Jun and Liu, Runnan and Ou, Wenwu},
  journal={arXiv preprint arXiv:2601.02955},
  year={2026}
}

@article{wang2026onelive,
  title={OneLive: Dynamically Unified Generative Framework for Live-Streaming Recommendation},
  author={Wang, Shen and Huang, Yusheng and Yang, Ruochen and Wen, Shuang and Xu, Pengbo and Cao, Jiangxia and Liu, Yueyang and Cai, Kuo and Guo, Chengcheng and Wang, Shiyao and others},
  journal={arXiv preprint arXiv:2602.08612},
  year={2026}
}

@inproceedings{cao2025pantheon,
  title={Pantheon: Personalized multi-objective ensemble sort via iterative pareto policy optimization},
  author={Cao, Jiangxia and Xu, Pengbo and Cheng, Yin and Guo, Kaiwei and Tang, Jian and Wang, Shijun and Leng, Dewei and Yang, Shuang and Liu, Zhaojie and Niu, Yanan and others},
  booktitle={Proceedings of the 34th ACM International Conference on Information and Knowledge Management},
  pages={5575--5582},
  year={2025}
}

@article{liang2024ensure,
  title={Ensure timeliness and accuracy: A novel sliding window data stream paradigm for live streaming recommendation},
  author={Liang, Fengqi and Zheng, Baigong and Zhao, Liqin and Zhou, Guorui and Wang, Qian and Niu, Yanan},
  journal={arXiv preprint arXiv:2402.14399},
  year={2024}
}


\end{document}